\documentclass[preprint,12pt]{aastex}

\usepackage{graphicx}

\begin{document}

\title{Detecting Transits of Planetary Companions to Giant Stars}

\author{R.J.~Assef\footnote{e-mail: {\tt{rjassef@astronomy.ohio-state.edu}}}
, B.S.~Gaudi and K.Z.~Stanek}

\affil{Department of Astronomy, The Ohio State University, 140 W.\
  18th Ave., Columbus, OH 43210}

\begin{abstract}

Of the approximately 350 extrasolar planets currently known, of order
10\% orbit evolved stars with radii $R_* \gtrsim 2.5 R_{\odot}$.
These planets are of particular interest because they tend to orbit
more massive hosts, and have been subjected to variable stellar
insolation over their recent histories as their primaries evolved off
the main sequence.  Unfortunately, we have limited information about
the physical properties of these planets, as they were all detected by
the radial velocity method and none have been observed to
transit. Here we evaluate the prospects for detecting transits of
planetary companions to giant stars.  We show that several of the
known systems have a priori transit probabilities of $\gtrsim 10\%$,
and about one transiting system is expected for the sample of host
stars with $R_* \ge 2.5 R_{\odot}$.  Although the transits are
expected to have very small amplitudes ($\sim \rm few \times 10^{-4}$)
and long durations ($\gtrsim 50 \rm hrs$), we argue that the
difficulty with detecting these signals in broadband light is one of
systematic errors and practicality rather than photon noise, even for
modest aperture ($\sim 1~{\rm m}$) telescopes. We propose a novel
method that may overcome these difficulties, which uses narrow-band
measurements to isolate the thin ring of chromospheric emission
expected at the limb of giant stars. The transit signals in these
narrow bands are expected to be larger in magnitude and briefer in
duration than in broad-band emission, and thus alleviating many of the
difficulties with transit detection in broad-band emission. Finally,
we point out that it may be possible to discover planetary companions
to giant stars using Kepler, provided that a sufficient number of such
targets are monitored.

\end{abstract}


\section{Introduction}\label{sec:intro}

Since the first extrasolar planet orbiting a main-sequence star was
detected by \citet{mayor95}, the number of known planets has increased
very rapidly.  To date we know of about 300 exoplanets in the local
solar neighborhood. Most of these were detected via the radial
velocity (RV) variations they induce on their host star, and the
typical hosts of these planets are F, G and K main sequence stars.
This is illustrated in Figure 1, which shows a color-magnitude diagram
of stars with planets originally detected with RV.

Among this sample of planets are a class of Jovian-mass
($\sim 0.1M_J$ to a few $M_J$) planets with periods of $\la 10~{\rm
days}$.  These ``Hot Jupiters'' were quite unexpected in the standard
core-accretion planet formation model, which was developed to explain
the solar system, and set the expectations for extrasolar planets
\citep{lissauer87}. The environmental conditions and evolutionary
history of these planets are quite distinct from that of the giant
planets in our solar system.  In particular, the energy budget of hot
Jupiters is dominated by the immense stellar isolation flux they
receive, which is a factor of $\sim 10^4$ higher than that for
Jupiter, and a factor of $\sim 10^4$ higher than the cooling flux of
an isolated planet after a few gigayears.  This large stellar
insolation has a profound effect on the properties of these planets.
In particular, their outer atmospheres develop a deep, isothermal
layer above the convective zone and below the photosphere, which slows
the rate at which the planet cools.  If these hot Jupiters migrated to
their present, high stellar insolation environment on a timescale that
is short compared to their cooling time in isolation of $\sim
10^6-10^7~{\rm yrs}$, then their rate of cooling and hence contraction
will be significantly slowed. As a result, hot Jupiters are predicted
to be significantly larger than Jupiter at their current ages of a few
Gyr.  At fixed mass and composition, their radii are predicted to be
correlated with their equilibrium temperatures ($T_{eq}$). Jovian-mass
planets orbiting solar-type stars beyond $\sim 0.1~{\rm AU}$ are hence
predicted to have smaller radii than hot Jupiters, although still
slightly inflated with respect to planets with $a\ga 1~{\rm AU}$
\citep{fortney07}.  In contrast, super-Jupiter planets more massive
than a few $M_J$ will have higher surface gravities, are expected to
contract more quickly than Jovian-mass planets, and are more nearly
purely degenerate.  Therefore, their radii are expected to be much
less sensitive to their age (for ages $\ga 10^8~{\rm yrs}$),
composition, or the amount of stellar insolation they receive
\citep{guillot05}.

Transiting planets offer the possibility of precision tests of these
theoretical predictions.  With a combination of radial velocity
measurements, stellar spectra, and precise photometry of the planetary
transit, it is possible to infer the radii, mass, and equilibrium
temperatures of transiting planets to an accuracy of a few percent
(e.g., \citealt{torres08}).  These measurements can be directly compared
with detailed theoretical model predictions
\citep{burrows00,hubbard01,burrows03,bodenheimer03,laughlin05},
although such comparisons must allow for unknown or poorly constrained
parameters, such as the mass in heavy elements, and the age of the
star.  These comparisons generally verify the most basic prediction of
these models, namely that the radii of hot Jupiters should be inflated
with respect to similar planets in isolation
\citep{burrows00,laughlin05}.  However, there are significant
discrepancies as well, and in particular a significant fraction of the
known transiting planets have radii that are far too large to be
explained by the simplest models, even including the effects of
stellar irradiation.  This is a long-standing problem \citep{burrows03},
and although many solutions have been proposed
\citep{bodenheimer03,guillot02,winn05,burrows07,mardling07,jackson08},
no consensus has yet emerged as to the correct explanation.

Precise radii of planets over a wider range of physical and
environmental conditions would, in principle, provide more decisive
tests of these models, and may even shed light on the origin of the
anomalously large radii.  Unfortunately, photometric transit surveys,
which have discovered the majority of transiting planet systems, are
strongly biased toward short-period systems \citep{gaudi05}.
Furthermore, these surveys are sensitive to the transiting planets
orbiting main-sequence stars with apparent magnitudes of $V=10-13$,
which are dominated by F and G dwarfs \citep{pepper03}.  As a result,
the majority of known transiting planets occupy a fairly narrow region
of parameter space, namely semimajor axes of $\la 0.1~{\rm AU}$ and
main-sequence hosts with $0.7M_\odot \la M \la 1.5 M_\odot$.  Radial
velocity surveys are less strongly biased with respect to semi-major
axis than transit surveys, and indeed the transiting planet with the
lowest-mass host, and the transiting planet with the largest
semi-major axis, were both first discovered by RV
\citep{butler04,fischer07}, and subsequently shown to be transiting
\citep{gillon07,barbieri07}.  Follow-up searches for additional
transiting systems among the sample of known planets detected by RV
\citep{seagroves03} may uncover some with long periods and
thus considerably expand the parameter space over which giant planet
models are tested.  Space-based transit missions are also less biased
with respect to semi-major axis, and should detect many long-period
transiting planet systems \citep{beatty08,yee08}.

While the majority of the nearby exoplanet hosts are main-sequence FGK
stars with masses of $0.7M_\odot \la M \la 1.5M_\odot$ and radii of
$0.7R_\odot \la R_* \la 2.5~R_\odot$, roughly $\sim 10\%$ are evolved
stars with $R_*\ga 2.5~R_\odot$.  These giant stars are interesting
targets for planet searches because they tend to have main-sequence
progenitors that are earlier spectral type and so more massive than
exoplanet hosts on the main-sequence.  As both the mass of the
protoplanetary disk and the dynamical time at the location of the snow
line are expected to depend on the mass of the primary, it is a
generic prediction that the mass and/or frequency of giant planets
should depend on the mass of the primary \citep{ida05,kennedy08}.
Searching for planets orbiting main-sequence stars much earlier than
spectral type F5 $(M\ga 1.5~M_\odot$) is challenging due to the fact
that these stars are hot and thus have few spectral lines in the
visible.  Furthermore, these stars tend to be fast rotators, and thus
what spectral lines are available tend to be broad.  Although there
are RV surveys that target these stars \citep{galland05}, they are
generally only sensitive to massive, super-Jupiters and brown
dwarf companions \citep{galland06}.  When these stars evolve off the
main-sequence, however, they expand and cool, thus making them much
more amenable to RV surveys \citep{sato03,johnson07a}.

To date, $\sim 25$ planetary companions to nearby stars with
$M>1.5~M_\odot$ have been discovered. Four correspond to planets
orbiting main-sequence A stars detected by direct imaging, namely
Fomahault b \citep{kalas08} and HR 8799 b, c and d \citep{marois08},
while the rest of the hosts are exclusively evolved stars with $R_*\ga
2.5 R_\odot$. Planetary companions to giant stars offer the
possibility of testing models of the structure and atmosphere of giant
planets for a relatively unique set of physical parameters and under a
relatively unique set of environmental conditions.  Consider, for
example, a $M_p\sim 3~M_J$ companion with a semimajor axis of $a\simeq
1~{\rm AU}$ orbiting an intermediate mass main-sequence A star with $M
\sim 2.0~M_\odot$.  If we assume that this planet formed {\it in
situ}, or migrated to this location on a timescale less than its
cooling time of $\la 10^{7}~{\rm years}$, it would have cooled in the
presence of a stellar insolation flux that is a factor of $\sim 40$
times smaller than a typical hot Jupiter companion to a $\sim M_\odot$
star.  Since the cooling time of the planet is significantly shorter
than the main-sequence lifetime of the host, it is expected that the
planet will cool and contract to a radius $\sim 1.1~R_J$
\citep{fortney07} before the host star leaves the main sequence.
Figure \ref{fg:evolve} shows the evolution of such a host after it
leaves the main sequence, according to the Padova stellar evolution
tracks from \citet{salasnich00}.  The luminosity at first decreases
slightly as it approaches the base of the giant branch, and then
begins to rise as the star ascends the giant branch.  Over the next
$\sim 20~{\rm Myr}$, the luminosity of the host increases by a factor
of $\sim 10$, while the radius increases by a factor of $\sim 5$.
Thus the insolation flux of the planetary companion increases by
factor of $\sim 10$, corresponding to an increase in the equilibrium
temperature of the planet of $\sim 400~{\rm K}$\footnote{Here we have
assumed, for simplicity, that all of the incident stellar radiation on
the planet is absorbed and re-emitted isotropically, and that these
characteristics, as well as the planet semimajor axis, do not evolve
with time.}. At the tip of the giant branch, the planet will have
$T_{eq}\sim 1000~{\rm K}$, similar to that of hot-Jupiter companions
to solar-type stars with $a\sim 0.1~{\rm AU}$.

As a result of the change in insolation, the atmospheric scale height
$H$ of the planet will increase on a thermal timescale of $\sim 10^2 -
10^4$ years \citep{showman08}, which is essentially instantaneous
compared to the timescale of changes in the insolation flux.  Note
that the bulk of the planet is not expected to be affected, as the
planet has already cooled, and the external heating cannot go against
the entropy gradient \citep{burrows00}.  Thus it is expected that the
changes in the stellar insolation will only affect the outer scale
height $H$ of the atmosphere, and thus the overall change in the
planet radius will be due only to the changes in $H$.  The scale
height $H$ will correlate linearly with equilibrium temperature for a
simple, isothermal atmosphere, such that
$$
 H=\frac{kT_{eq}}{\mu m_p g},
$$ where $\mu$ is the mean molecular weight, $g$ is the surface
gravity, and $m_p$ is the proton mass.  Accounting for the transit
radius effect (e.g., \citealt{burrows07}), changes in $H$ will then
result in changes in the apparent radius of a transiting planet of
$$
\frac{\Delta R_p}{R_p} \simeq \frac{H}{2R_p}\ln{\frac{2\pi R_p}{H}}.
$$ For this particular case, it is expected that the radius of the
planet will only increase by $\sim 0.1\%$ relative to its radius when
the host was on the main-sequence (Fig.\ \ref{fg:evolve}). This is
unlikely to be detectable.  However, for planets with lower surface
gravities, the effect will be larger and potentially detectable.

The basic prediction is therefore that the radii of the known
planetary companions to giant stars will have been largely unaffected
by the post main-sequence evolution of their hosts.  Nevertheless, it
would be quite interesting to test this prediction.  Furthermore, the
existence of the heretofore unexplained population of `bloated
planets' with radii larger than predicted by models suggests that this
prediction may not be entirely robust.  In particular, if, as
suggested by \citet{guillot02}, the bloated radii can be explained by
a small fraction of the stellar insolation energy being deposited deep
in the interiors of the planets through winds in the atmospheres, then
the large changes in the stellar isolation flux experienced by
planetary companions to evolved stellar hosts could lead to
significant changes in their radii. Since the transit depth is
proportional to the square of the radius of the planet, this would
have important implications for their detectability. Other observable
effects of the change in stellar insolation are also possible, such as
an increase in the atmospheric evaporation rate
\citep{baraffe04,hubbard07}.

Thus it would be of considerable interest to detect transiting
planetary companions to evolved stars, in order to measure their radii
and so test models of the structure, evolution, and atmospheres of
these planets.  Unfortunately, all of the planetary companions to
giant stars have been found by RV surveys and none of them have been
observed in transit, so our knowledge of them is very limited. In this
paper we discuss the possibility of detecting transits of the known
planets orbiting giant stars. In \S~\ref{sec:trans_prob} we discuss
the probability of seeing one of these planets transiting its parent
star, while in \S~\ref{sec:transits} we address the difficulties and
prospects of detecting them. In \S\S~\ref{sec:limb_bright} and
\ref{sec:trans_HK} we present a novel technique for observing
planetary transits in giant stars by using narrow-band filters that
could potentially overcome all the difficulties inherent in their
broad-band signals.

\section{Transit Probability}\label{sec:trans_prob}

\begin{figure}
  \begin{center}
    \plotone{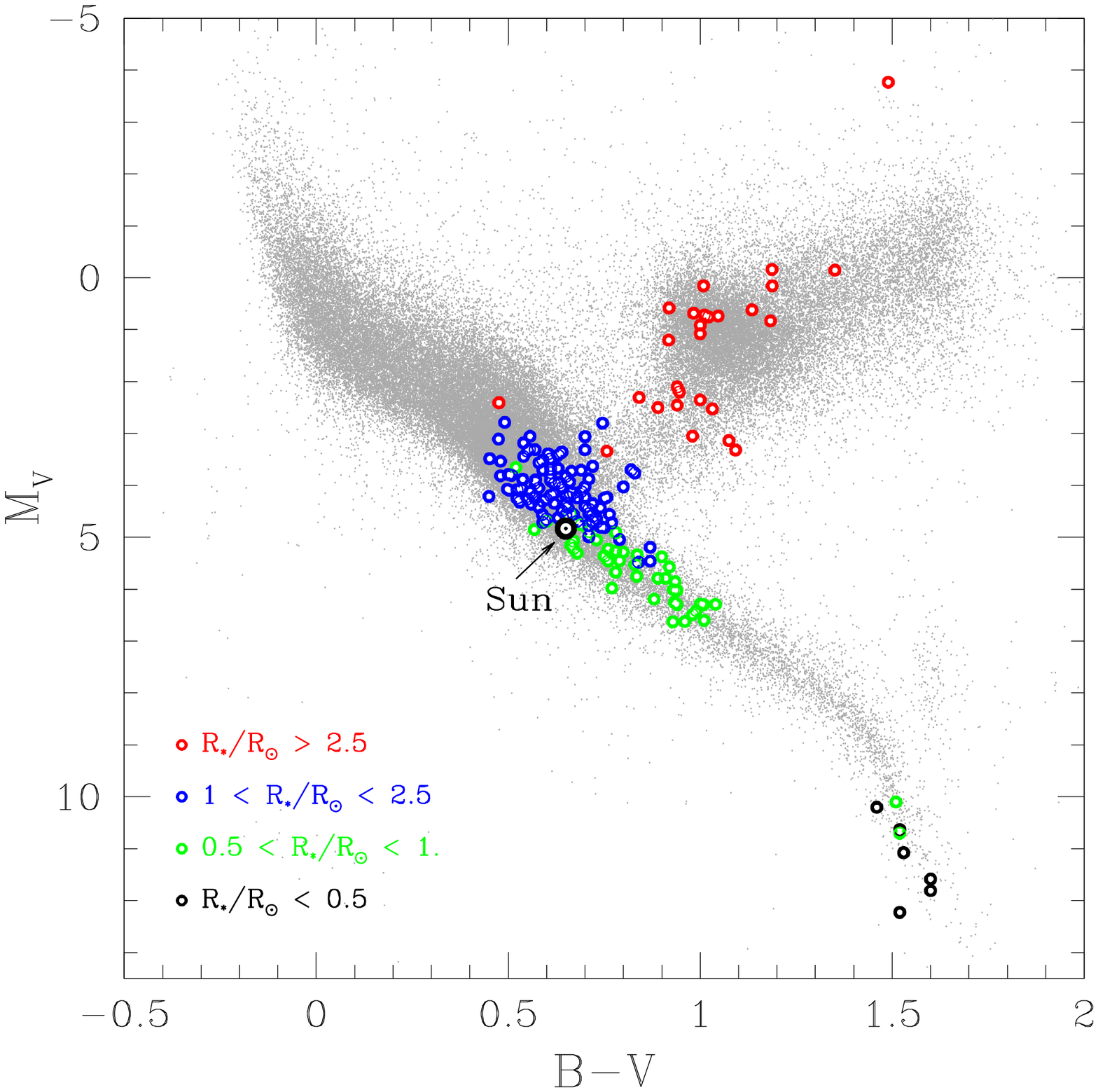}
    \caption{Color-magnitude diagram of the stars that were found by
    RV surveys to host exoplanets (\textit{solid circles}). Stars are
    color coded by their radius: $R_*<0.5R_{\odot}$ ({\textit{black}}),
    $0.5<R_*<1. R_{\odot}$ ({\textit{green}}), $1.<R_*<2.5 R_{\odot}$
    ({\textit{blue}}) and $R_* > 2.5 R_{\odot}$
    ({\textit{red}}). When a value for the radius was not listed
    by {\it The Extrasolar Planets Encyclopedia} we estimated it from
    the \citet{vanbelle99} color--angular size relations combined with
    the Hipparcos mission parallax measurements. For reference, we
    also show all the Hipparcos sources (\textit{grey dots}).}
    \label{fg:cmd}
  \end{center}
\end{figure}

\begin{figure}
  \begin{center}
    \plotone{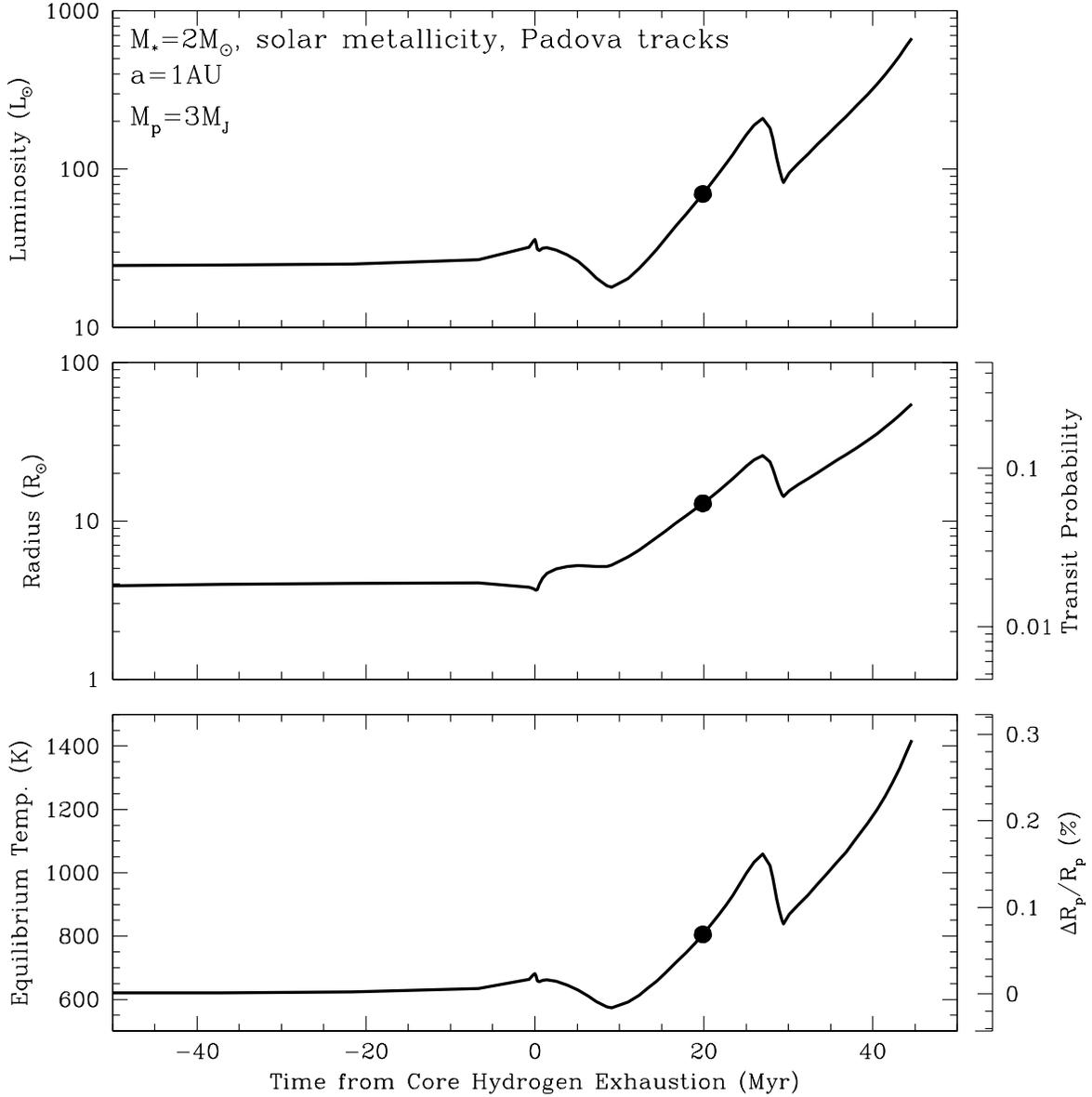}
    \caption{Evolution of a $2 M_\odot$ star as a function of time
      in Myr relative to the time when the hydrogen in its core is
      exhausted, i.e.\ the end of its main-sequence lifetime, from
      the evolutionary tracks of \citet{salasnich00}.  Top Panel:
      Luminosity as a function of time.  Middle Panel: Radius as a
      function of time.  The extreme right axis shows the transit
      probability as a function of time for a planet at $a=1~{\rm
      AU}$.  Bottom Panel: The evolution of the equilibrium
      temperature of a planet at $1~{\rm AU}$ as a function of time.
      Here we have assumed an albedo of $A_B=0$ and complete
      redistribution of heat.  The extreme right axis shows the
      corresponding change in the radius of a planet with
      $M_p=3~M_J$, relative to its radius when the host
      is on the main-sequence.  In all three panels, the dot shows the
      parameters of a star with a post-main sequence age of $\sim
      20~{\rm Myr}$: $L \simeq 70~L_\odot$, $R_*\simeq 14 R_\odot$, and
      $T_{eq}\simeq 800~{\rm K}$.  The stellar and planetary
      parameters of this system approximates those of the known
      planet/star system HD 173416 \citep{liu09}.}
    \label{fg:evolve}
  \end{center}
\end{figure}

As summarized previously, several tens of planets orbiting giant stars
have been discovered by radial velocity surveys, but to date none has
been observed to transit its parent star. Indeed, to the best of our
knowledge, no attempt has been made to search for transits among the
known planetary companions to giants. This is likely due to the
dauntingly small expected signal, and the difficulties in detecting
such small signals from ground-based observations. We discuss these
issues further in the next section. Nevertheless, there is a
significant probability that one of these planets should be transiting
as seen from Earth. The a priori probability that a planet will
transit is given by \citep[e.g.][]{seagroves03}
\begin{equation}\label{eq:trans_prob}
P_{\rm tr}\ =\ 0.0045\ \left(\frac{1 \rm AU}{a}\right)\
\left(\frac{R_*+R_{p}}{R_{\odot}}\right)\ \left[\frac{1+e
\cos(\frac{\pi}{2}-\varpi)}{1-e^2}\right],
\end{equation}
\noindent where $a$ is the semi-major axis of the orbit, $R_*$ and
$R_{p}$ are respectively the star's and planet's radius, $e$ the
eccentricity of the orbit and $\varpi$ the longitude of
periastron. The probability of transit is directly proportional to the
star's radius but inversely proportional to the semi-major axis of the
planet.  Thus it is not immediately obvious whether the probability is
higher or lower for the planetary companions to giant stars as
compared to typical transiting hot Jupiters, as the larger radii of
the giant hosts compensate for the larger typical separation of their
planetary companions \cite[see][]{sato08}. To estimate the number of
known planets around giant stars we would expect to observe in
transit, we obtain the public list of known planets found by RV
surveys as compiled by {\it{The Extrasolar Planets
Encyclopedia}}\footnote{http://exoplanet.eu/ as of October 10$^{\rm
th}$, 2008}. We first consider all types of stellar hosts. To estimate
the transit probability for each system, we obtain $a$, $e$, $\varpi$,
$R_*$ and $R_{p}$ from this list whenever possible. If $R_{p}$ is not
listed, we assume the planet to be of Jupiter size. If $R_*$ is not
listed, we estimate it from the angular size -- color relations of
\citet{vanbelle99} and parallax from Hipparcos \citep{perryman97}. The
angular size -- color relation is different for main sequence and
evolved stars, which poses a problem since we generally do not have a
priori information of the luminosity class of the objects.  We assume
that a given star is on the main sequence unless the determined radius
is larger than $1 R_{\odot}$ (which is roughly the local turn off
radius), in which case we recalculate it with the relation for evolved
stars. Recently, these calibrations have been revised by
\citet{vanbelle09} but we have not updated our calculations as the
changes are not too significant for the relevant objects. For all
stars we obtain $V$ band photometry from
SIMBAD\footnote{http://simbad.u-strasbg.fr/simbad/}, and $K$ band
photometry from the Two Micron All Sky Survey (2MASS) All-Sky Catalog
of Point Sources \citep{cutri03} via the Vizier
service\footnote{http://vizier.u-strasbg.fr/}. Five exceptions to this
process must be noted. (1) We eliminate TW Hya b from our sample
because its existence has recently been questioned by
\citet{huelamo08}. (2) We correct the radius of the star HD13189 to
that measured by \citet{baines08}, from the value listed by {\it{The
Extrasolar Planets Encyclopedia}}, which is 100 times larger.  (3) We
correct the V-band magnitude of the M Dwarf GJ317, as SIMBAD lists it
to be 1 magnitude fainter than what is commonly quoted
\citep{johnson07}. (4) The B- and V-band magnitudes of HD330075 are
corrected to the values reported by \citet{pepe04}, which are about
0.3 magnitudes fainter in both bands than the values listed by
SIMBAD. (5) For HD47536b, no value for the semi-major axis is listed
in the {\it The Extrasolar Planets Encyclopedia}, and so we estimate
$a$=1.5AU based on the mass of the primary determined by
\citet{setiawan08} and the orbital period determined by
\citet{setiawan03}. The color magnitude diagram of all the host stars
in our sample is shown in Figure \ref{fg:cmd}, together with all the
sources in the Hipparcos catalog. It is clear that most exoplanet
hosts are main-sequence stars with $0.4 \la B-V \la 1$.  About $13\%$
of hosts are evolved stars with $R_*\ga 2.5R_\odot$.  The majority of
these are red clump stars.  The largest exoplanet host star is HD13189
with a radius $\sim 50~R_\odot$, which has a $M_p\simeq 14~M_J$
companion at a separation of $a\simeq 1.85$ AU.

\begin{figure}
  \begin{center}
    \plotone{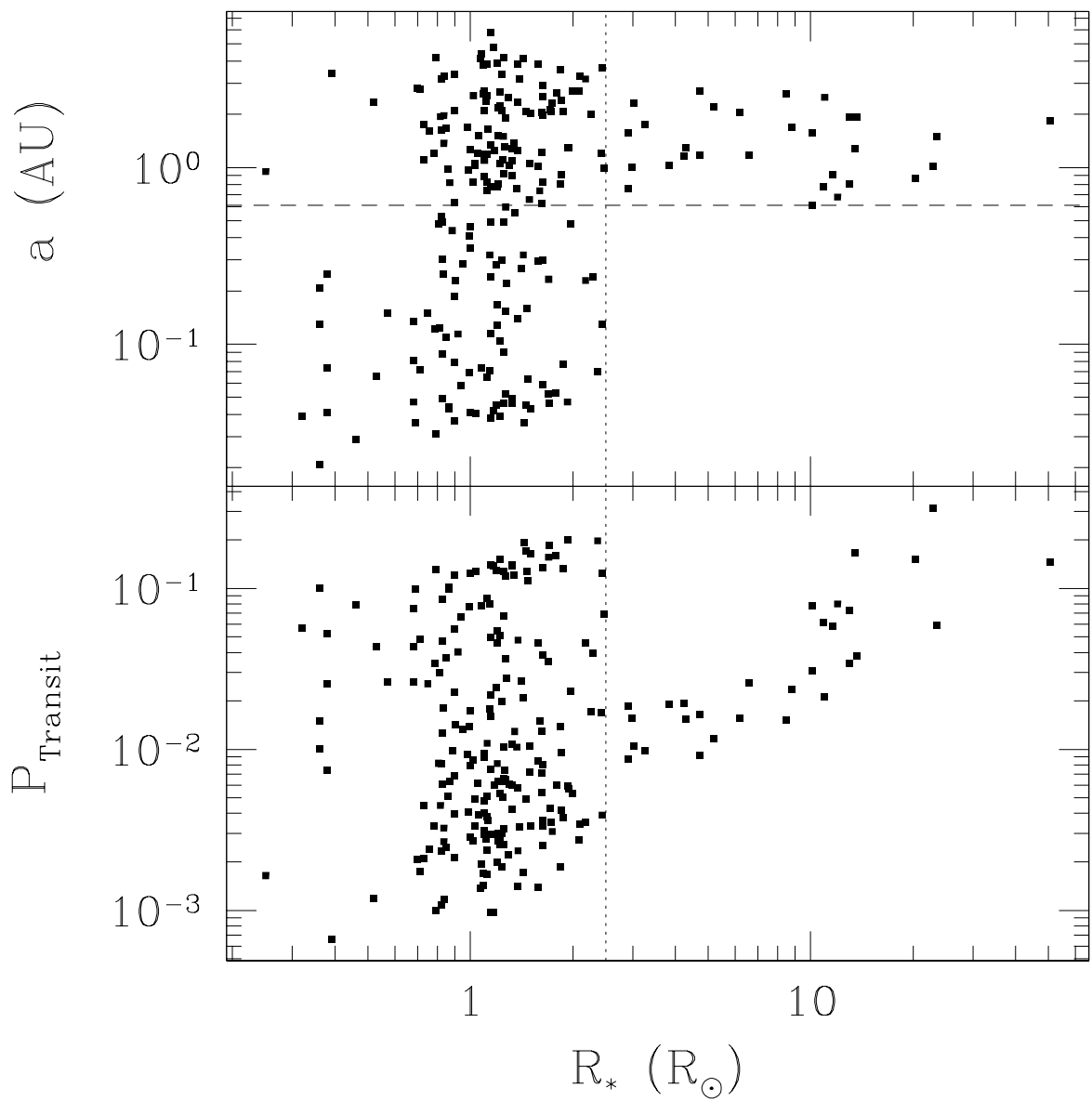}
    \caption{Semi-major axis ({\textit{top}}) and transit probability
    ({\textit{bottom}}) as a function of stellar radius for stars with
    exoplanets detected by RV, as culled from {\it The Extrasolar
    Planets Encyclopedia}.  The dashed line in the top panel shows the
    minimum semi-major axis for planets around giant stars determined
    by \citet{sato08}. The vertical dotted line indicates our adopted
    radius cut of $2.5 R_{\odot}$ for selecting giant stars.Note that
    our sample is composed 29 planets orbiting the 27 giant stars
    shown on Figure \ref{fg:cmd}. The two multiple planetary systems
    correspond to HD102272 and HD60532.}
    \label{fg:prob}
  \end{center}
\end{figure}

Figure \ref{fg:prob} shows the transit probability and semi-major axis
as a function of the radius of the host star for the systems in our
sample. For main sequence stars, $\log{a}$ and thus the transit
probabilities appear to be roughly distributed uniformly for $a\ga
0.04$ AU. For evolved hosts with $R_*\ga 2.5 R_{\odot}$, however, no
system has a semi-major axis smaller than $0.61AU$
\citep{niedzielski08,sato08}. There are four systems that have a
priori transit probabilities greater than 10\%.  These are 4UMa b
($P_{\rm tr}\simeq 0.15$), HD122430 b ($P_{\rm tr}\simeq 0.32$),
HD13189 b ($P_{\rm tr}\simeq 0.15$) and HIP75458 b ($P_{\rm tr}\simeq
0.17$).  These may be the best targets to search for transiting
companions, although they are also expected to have the smallest
transit signals, with depths of $2.55\times 10^{-5}, 2.01\times
10^{-5}, 4.16\times 10^{-6}$, and $5.80\times 10^{-5}$, respectively,
for a companion with $R_p=R_J$.

\begin{figure}
  \begin{center}
    \plotone{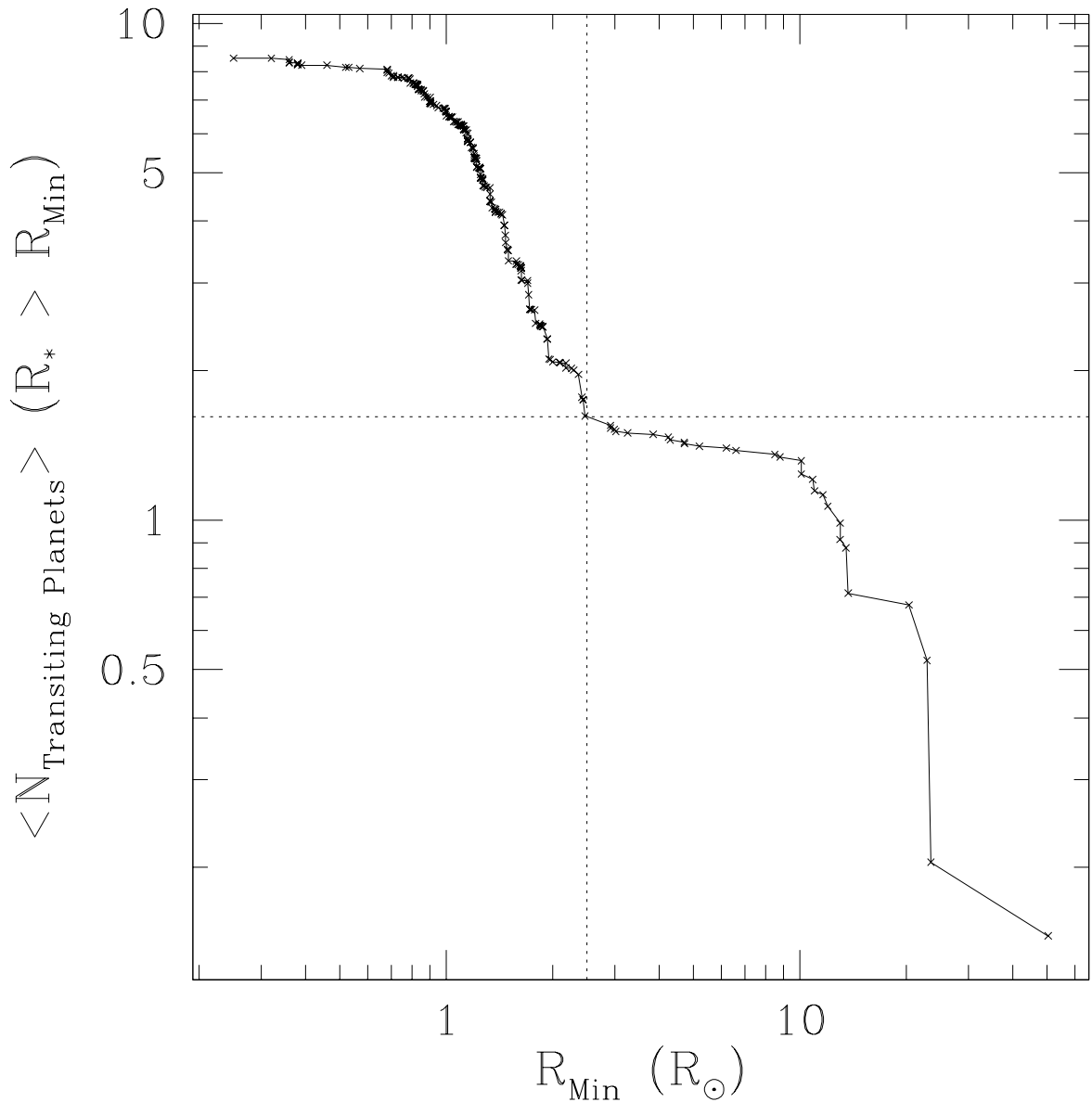}
    \caption{Expected number of transiting planets as a function of
    the minimum radius of the host star for exoplanets found by RV
    surveys. Each cross shows expected number of transiting systems,
    calculated by successively eliminating the smallest host star from
    total sample of RV-detected exoplanets culled from the {\it The
    Extrasolar Planets Encyclopedia} list. The vertical dashed line
    shows our adopted radius cut of $2.5 R_{\odot}$ for giant stars,
    while the horizontal dashed line shows the expected number of
    transiting planets orbiting hosts stars above this radius
    threshold linearly interpolated between the two adjacent points.}
    \label{fg:n_expect}
  \end{center}
\end{figure}

Figure \ref{fg:n_expect} shows the number of planets detected by RV
surveys that we expect to be transiting, as a function of the minimum
stellar radius of the hosts. If we consider only planets orbiting
stars larger than $2.5R_{\odot}$, we expect about one to be
transiting.

\section{Detecting Planets Around Giant Stars}\label{sec:transits}

In the previous section we demonstrated that we expect of order one
transiting companion among the exoplanet host stars with radii larger
than $2.5 R_{\odot}$. Despite this, to the best of our knowledge, no
follow up observations are being conducted to search for these
transits.

Nominally, the signal to noise ratio ($S/N$) for the transit of a
planet in front of a giant star can be very large. Consider, as a
fiducial example, a planet with $R_p=R_J$ orbiting a red giant star
with mass $M_*=1M_{\odot}$ and radius $R_*=5R_{\odot}$. These values
are typical of red giants in the solar vicinity. Recently,
\citet{sato08} found that planetary companions to giant stars with
intermediate-mass ($1.7 < M < 3.9 M_{\odot}$) main-sequence
progenitors appear to have a minimum semi-major-axis of 0.68AU.
Since, at fixed $R_*$, the closest planets will have the highest
transit probability, we will therefore assume for our fiducial case
$a=0.7\rm AU$. In the absence of correlated noise, the $S/N$ of a
transit is
\begin{equation}\label{eq:sn_simple}
\frac{S}{N}\ \sim\  n^{1/2}\ \frac{\delta}{\sigma},
\end{equation}
\noindent where $n$ is the number of data points during transit,
$\delta$ is the fractional transit depth, and $\sigma$ is the
fractional photometric precision. If photons are collected at a rate
$\Gamma$, and the transit duration is $t_T$, then
\begin{equation}\label{eq:sn_gamma_t}
\frac{S}{N}\ \sim\ \sqrt{\Gamma t_T}\ \delta. 
\end{equation}
For our example case 
\begin{equation}\label{eq:delta_example}
\delta\ =\ \left(\frac{R_p}{R_*}\right)^{2} \approx 4\times
10^{-4}\ \left(\frac{R_p}{R_J}\right)^2\ \left(\frac{R_*}{5
R_{\odot}}\right)^{-2},
\end{equation}
\noindent where $R_J$ is the radius of Jupiter, and the transit
duration is
\begin{equation}\label{eq:t_T_example}
t_T\ \approx\ 54\ {\rm hours}\ \left(\frac{a}{0.7 {\rm AU}}\right)^{1/2}\
\left(\frac{R_*}{5 R_{\odot}}\right)\
\left(\frac{M}{M_{\odot}}\right)^{-1/2},
\end{equation}
where we have assumed a circular orbit for simplicity.  The average
$V$-magnitude of the exoplanet hosts with $R_* > 2.5 R_{\odot}$ is $V =
5.5$. If we assume a telescope with a 1m aperture with and overall
efficiency of 50\% (i.e., we detect 50\% of the photons from the
star), the $S/N$ of our example case would be $\sim 870$ per transit.

Even though this nominal value for the $S/N$ ratio is extremely large,
it must be taken into account that the transit duration $t_T$ is long
enough that such a transit cannot be observed from a single facility
unless it is located in space. In practice, a ground-based
telescope would only be able to observe a small part of the transit,
significantly degrading the signal to noise ratio. If neither ingress
or egress are observed, the photometry is unlikely to be sufficiently
stable from night-to-night to identify transits of such small
depths.

Even if the difficulties posed by the long transit duration could be
overcome by, for example, a network of telescopes around the world
with properly accounted photometric offsets, the transit might still
be very hard to detect because of correlated errors in the relative
photometry, commonly known as ``red noise.'' It is unclear what
exactly causes this noise, but it is most likely a combination of
effects including seeing variations, airmass changes and flatfielding
errors. These correlations introduce coherent modulations in the light
curves that can corrupt the transit signal (see \citealt{pont06} for a
detailed discussion). In current photometric transit surveys, the
typical level of the red noise is of the order of $10^{-3}$ magnitudes
and is thought to be correlated in timescales of a few hours
\citep{gould06,pont07}. Since the transit signal of a Jupiter-sized
planet around a giant star is significantly smaller than this (for our
fiducial case it is $\sim 4\times 10^{-4}$ magnitudes) current transit
surveys are unable to detect transiting Jupiter-sized planets around
giant stars, although it should be kept in mind that it is generally
not known how red noise behaves on the timescales relevant for this
problem. Errors are known to be less correlated between consecutive
nights, therefore it might be the case that the very long duration of
planetary transits of giant stars mitigates the effect of the red
noise.

Targeted photometric follow-up of transits in bright stars can achieve
very low levels of systematic noise even when the observations are
taken from the ground, as controlling the possible causes of red noise
is much simpler when all efforts are concentrated on a single
system. For example, \citet{winn07} performed ground-based photometric
follow-up observations of the planetary system TrES-1 \citep{alonso04}
and were able to reduce the correlated errors in short timescale
observations ($\sim 3$ hours) to a level smaller than $10^{-4}$
magnitudes. See also \citet{johnson09} and \cite{winn09}.
\citet{nutz08} observed transits of HD149026b from space with the
8$\mu$m channel of IRAC on Spitzer and controlled the systematic
errors to a very similar level on longer timescales ($\sim 7$
hours). While the regimes in which these observations have been made
do not necessary apply to planetary transits in giant stars, their
precision is very encouraging and suggest that, in the near future,
transit follow-up observations of known planets around giants could
achieve the photometric precision and control of systematics necessary
to detect transiting Jupiter-size companions. Nevertheless, a note of
caution must be drawn, as one of the ways these surveys control
systematic errors is by removing long timescale photometric trends,
such as linear changes in the flux over a night or night-to-night
average flux variations, as these could be caused by, for example,
flatfielding errors. For close-in planets this does not represent a
problem, as the transit durations are short compared to the scales
over which these trends appear, but the transit durations for planets
around giant hosts are considerably longer, and thus their signal
could be interpreted as due to systematic errors in the photometry.

With their continuous photometry and better control of systematic
errors, space missions like {\it Kepler} \citep{kepler} and {\it
COROT} \citep{corot} could be very sensitive to Jupiter-sized
companions transiting in front of giant stars in their field-of-view,
and thus have the potential to discover such systems.  Indeed, the
depth of the transit of a $\sim R_J$ planet transiting in front of a
$\sim 10R_\odot$ star is quite comparable to the transit depth of an
Earth-sized planet transiting in front of a solar-type star, precisely
the signal that {\it Kepler} was designed to detect.  Furthermore, for
planets at $a\sim 1$ AU, the duration of the transit signal is $\sim
10$ times longer.  If we assume that red noise in Kepler is negligible
and the $S/N$ can be estimated by means of equation
(\ref{eq:sn_gamma_t}), taking the value of $\Gamma$ listed in Kepler's
official website\footnote{http://kepler.nasa.gov} yields $S/N \approx
35$ per transit for $R_*=5R_\odot$, $R_p=R_J$, $a=0.7$ AU, and $V=14$,
which is near the faint limit of {\it Kepler}.  The field of view of
{\it Kepler} is estimated to have 223000 stars brighter than $m_{\rm
v} = 14$ of which 136000 should be on the main sequence. Most of the
87000 other stars should be red giants, but currently most of them are
set to be eliminated before the data is downloaded from the
satellite. If only 1000 of them were downloaded, assuming a radius of
$R_*=5 R_{\odot}$ for these stars, and that 5\% of them host a
Jupiter-sized planet at $a=1$ AU, the expected number of transiting
systems is approximately 1.

All the estimates we have derived here assume that the intrinsic
variability on the time-scales relevant to planetary transits is small
compared to the depth of the transits. While this is true and very
well studied for F, G and K dwarfs, little is known about the
variability of giant stars on the time-scales of tens of
hours. \citet{baliunas81} studied the short time-scale variability of
chromospheric CaII of 4 giants ($\alpha$ Boo, $\alpha$ Tau, $\alpha$
Aur and $\lambda$ And) using high resolution spectra, and for two of
them ($\alpha$ Tau and $\lambda$ And) also using the HKP2
spectrophotometer on the Mount Wilson 60-inch telescope (see
\S~\ref{ssec:trans_detect} for a more detailed description of the
instrument). While no variability of the CaII H and K lines was found
in the high resolution spectra, the spectrophotometry shows some
significant quasi-regular variability of 10\% or less on timescales of
8--30 minutes for $\lambda$ And and of 25--30 minutes for $\alpha$
Tau. \citet{adelman01} studied the variability of red clump giants in
the Hipparcos database and found most of them to be constant within
0.03 mag. Several studies of this type of variability have been done
for Mira type variables with extremely discrepant
results. \citet{smak79}, \citet{maffei95} and \citet{delaverny98}
observed short time-scale broad-band variability of Mira variables in
the visible and near-IR. In particular, \citet{delaverny98} observed
51 variability events of 0.23 to 1.11 mag variations with time-scales
of 2 hours to 6 days in a sample of 39 Miras from the Hipparcos
catalog. In contrast, \citet{smith02} found no short time-scale (hours
to days) variability in the near- to mid-IR observations of 38 Mira
variables from the COBE DIRBE \citep{hauser98} database. Also,
\citet{wozniak04} failed to detect this effect in a sample of 485
Miras in the Galactic bulge based on $I$-band observations from the
OGLE-II experiment \citep{ogleII}. They derive an average upper limit
on the rate of irregular rapid-variability events of the type found by
\citet{delaverny98} of 1 per 26 years per star, in striking contrast
to the rate of 1 event per year per star implied by the
\citet{delaverny98} results. They identify three possible resolutions
to this apparent discrepancy: (1) the Mira variables in the Galactic
Bulge population differ from their counterparts in the solar
neighborhood, (2) the variability is in a part of the spectrum not
covered by the Cousins $I$-band used by OGLE-II, or (3) the short
time-scale variability events are much less frequent than suggested by
\citet{delaverny98}. Recently, \citet{aigrain09} studied the
noise properties of the {\it{COROT}} data from its first 14 months of
observations and determined that stars identified as likely red giants
exhibited a magnitude-independent noise scatter of 0.5 mmag on 2 hour
timescales, significantly higher than for stars identified as likely
dwarfs. They interpret this increased scatter as intrinsic photometric
variability. However, it is not clear how this amplitude of intrinsic
variability should be extrapolated to the 50 hr timescale relevant for
planetary transits. \citet{kallinger08} studied the Fourier spectrum
of 31 likely red giants in the {\it{COROT}} {\it{exofield}} and found
that these stars show a variability of 0.05--0.1 mmag in timescales of
50 hrs ($\sim 5\mu$Hz). This amplitude of variability is of the order
of the expected transit depths for giant stars and could significantly
hinder the ability to identify them. Without a consistent picture of
variability in red giants it is not possible to assess its effects on
the detection of planetary transits, although it is very encouraging
that several independent studies have found either null results or
amplitudes of variability that would not be disastrous for their
identification.

While it seems that in the near future detecting transits around giant
stars could be possible with current techniques, in the following
sections we show that it might also be possible to detect them by
using narrow band filters centered on the CaII H and K lines rather
than with the commonly used broad-band filters. Observations in these
bands may overcome many of the difficulties discussed above, as the
transit depths can be much deeper and the transit durations are much
shorter.

\section{Limb Brightening of Giant Stars}\label{sec:limb_bright}

The great majority of the flux from stars comes from their
photosphere, where gas is relatively cool compared to the inner
regions and elements leave imprints of absorption lines in the light
spectrum. Surrounding the photosphere, gas of significantly higher
temperature ($\sim 20,000$K) and lower density has been observed in
some stars. We call this atmospheric layer the chromosphere and,
because of its physical conditions, some resonant transitions that are
still optically thick in this region produce emission lines in the
spectrum. This means that while for most wavelengths the star will
appear to be brighter at the center and fainter towards the limb, at
the wavelengths of these transitions the radial surface brightness
distribution will have the opposite trend: it will appear as a faint
disc surrounded by a brighter limb.

Studying the emission lines coming from the chromosphere and their
possible variability can be crucial to understanding the heating
mechanisms of this region \citep[either magnetic, mechanic or a
combination of both; see e.g.][]{pasquini00} and processes of mass
loss in different stars \citep[e.g][]{dupree99}. A big drawback is
that most of these lines fall in the UV part of the spectrum,
rendering them unobservable from the ground. This makes the CaII
doublet (H and K lines) of particular interest, as it falls directly
in the optical region ($\lambda 3933$ \AA\, and $\lambda 3960$ \AA\,
for H and K respectively) and can exhibit very strong narrow
chromospheric emission in stars of types later than F0
\citep{shkolnik03}. In some cases, the emission line is double peaked,
exhibiting a centrally reversed profile \citep[e.g.][]{dupree99}.

While this emission is found in every star that has a chromosphere, it
is particularly important in red giant stars as their atmospheres are
significantly more extended than those of dwarfs. For example, the Sun
has been observed to show resonant line scattering in its limb over a
ring of a width of order $0.01 R_{\odot}$, while for red giants this
width can be 1--2 orders of magnitude larger \citep[see][and
references therein]{loeb95}.

Monitoring of the CaII H \& K emission in bright stars has been
routinely done for the last 30 years
\citep{wilson78,duncan91,wright04} and has been recently used as a
tool for estimating chromospheric activity and radial velocity
jitter in stars targeted by radial
velocity surveys for planets \citep{wright04}. In the next section we
describe a novel technique to observe planetary transits that takes
advantage of the limb brightened profile shown by red giant stars at
the wavelengths of CaII H \& K that may overcome many of the
difficulties with detecting transits using broad-band based
measurements.

\section{Transits in H \& K}\label{sec:trans_HK}

\subsection{Detecting Transits in H \& K}\label{ssec:trans_detect}

\begin{figure}
  \begin{center}
    \plotone{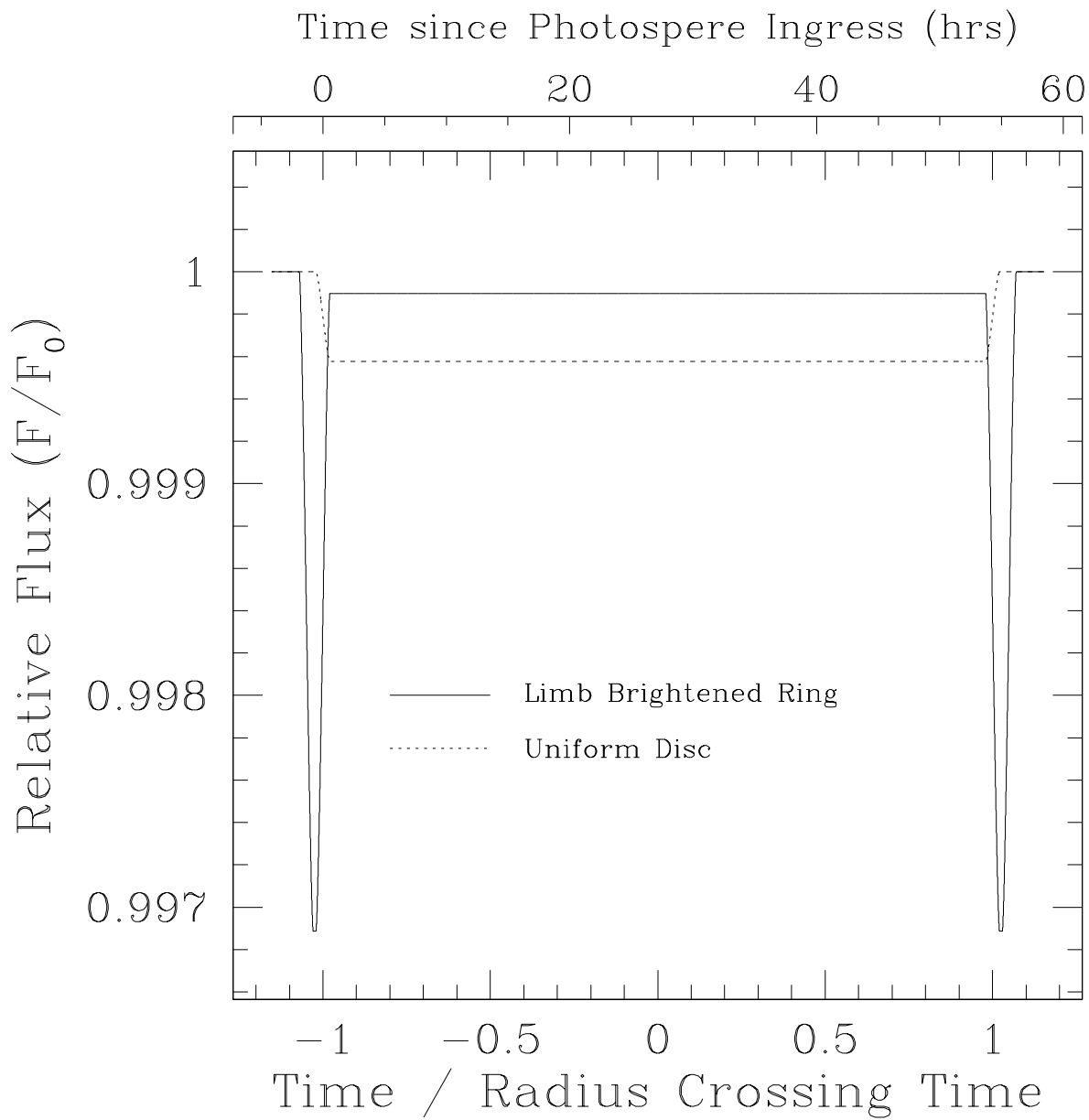}
    \caption{The solid line shows the light curve of a Jupiter-sized
    planet transiting a 5$R_{\odot}$ giant star, as observed through
    narrow bands centered in the core of the CaII H and K lines.  We
    have assumed a central transit. The dashed line shows the transit
    light curve for the same system but assuming a uniform surface
    brightness distribution, as would be observed through broad-band
    filters. For the narrow band observations, we have modeled the
    star as an inner disk surrounded by a ring of uniform brightness
    with an intensity 30 times larger than the inner disk. The width
    of the ring is 5\% of the radius of the star.}
    \label{fg:transit_comp} 
  \end{center}
\end{figure}

As described in the last section, if we were to observe a giant star
through narrow band filters centered on the CaII H \& K lines, we
would see a faint disc surrounded by a brighter ring, in sharp
contrast to the more familiar limb-darkened disc seen in broad-band
light. As the width of the CaII H \& K line emitting region is much
smaller than the radius of the star, transiting planets will block a
significantly larger fraction of the light than in the uniform disk
case. Figure \ref{fg:transit_comp} shows how the transit light curve
would appear in CaII H \& K for the same fiducial case discussed in
\S\ref{sec:transits}, a Jupiter-sized planet transiting a 5$R_{\odot}$
giant star, compared to the uniform disk model assumed before. The
depth of the transit is about 8 times larger for the former case and
is well above the red noise limit of current surveys when crossing the
brightened limb, while during the inner disk transit it is much
shallower. For simplicity, in this figure, and for the rest of the
section, we have assumed that the star can be modeled as an inner
uniform disk surrounded by a ring of width in units of the stellar
radius $R_*$ of $w = 0.05$ that has an intensity relative to the inner
disk intensity of $I = 30$. Note that $R_*$ refers to the photospheric
radius of the star rather than the chromospheric radius. The values of
$w$ and $I$ were taken to roughly reproduce the limb brightening model
of \citet{loeb95} for a 100$R_{\odot}$ red giant star. While this
model is probably not accurate for the smaller giant stars we consider
here, it is sufficient to provide an order of magnitude estimate of
the detectability of these types of transits. Below we will describe
how the $S/N$ depends on the exact value of these parameters. Proper
modeling of the chromospheric emission in giant stars involves
extensive non-LTE atmospheric calculations that go beyond the scope of
this work.

In this model, the chromospheric transit depth $\delta_R$ (i.e.\ the
transit depth when transiting the chromosphere but considering both
light components) can be easily described in two limiting cases.  When
the width of the ring is much smaller than the radius of the
transiting planet, the chromospheric transit depth is,
\begin{equation}\label{eq:delta_smallw}
\delta_R\ =\ \frac{p}{\pi}\ \left(\frac{2 I w + \pi p /2}{2 I w +
1}\right), \qquad (w \ll p),
\end{equation}
\noindent 
where $p\equiv R_p/R_*$. When the width of the ring is larger than the
planet's diameter, the chromospheric depth is
\begin{equation}\label{eq:delta_largew}
\delta_R\ =\ \frac{I p^2}{I w (2+w) + 1}, \qquad (w > 2p).
\end{equation}
Note that in the limit where the flux of the chromospheric ring is
much greater than that of the disc, equation (\ref{eq:delta_smallw})
converges to $\delta_R = p/\pi$, completely independent of the ring's
width.  Thus for $w\ll p$, the chromospheric transit depth is larger
than the broad-band (photospheric) depth by a factor of $\sim (\pi
p)^{-1}$.

It is worth noting that there exist instruments
specifically designed to observe the flux in the CaII H \& K
lines. They are commonly used in conjunction with radial velocity
surveys in order to monitor chromospheric activity in the target
stars. One such instrument is the HKP-2 \citep{vaughan78} at Mount
Wilson. To estimate the $S/N$ of the detection of one of these
transits, we will assume that the observations are carried out on the
$H$ and $K$ filters of the HKP-2 instrument. Both bands have a
triangular instrumental profile with a FWHM of approximately 1 \AA\, 
centered at the CaII H \& K lines respectively. Assuming that the
flux per unit wavelength $F_{\lambda}$ is the same as at the effective
wavelength of the Johnson's $B$-band, the signal to noise ratio is
given approximately by
\begin{equation}\label{eq:sn}
\left(\frac{S}{N}\right)^2\ =\ \left(\frac{2445}{\rm sec.}\right)\ e
N_T \left(\frac{A_T}{\rm cm^2}\right)\ 10^{-0.4 m_B}\ \int_{t_i}^{t_e}
\left(\frac{F(t)}{F_{\rm out}} - 1\right)^2 dt,
\end{equation}
\noindent where $N_T$ is the number of transits observed, $e$ is the
total efficiency of the observation (fraction of photons detected),
$A_T$ is the effective area of the telescope, $m_B$ is the $B$-band
magnitude of the star, $F(t)$ is the observed flux as a function of
time, $F_{\rm out}$ is the star's flux outside transit and $t_i$ and
$t_e$ are the times of first and last contact with the chromosphere
respectively. For the same fiducial case of \S\ref{sec:transits},
assuming $e=1/2$, we find $S/N \simeq 43$. If we eliminate the inner
disk part of the transit, since it is unlikely to be detected, the
$S/N$ drops almost negligibly to 42. Note that in equation
(\ref{eq:sn}) we have assumed that the $S/N$ is limited by the flux of
the source. The noise from the background should be negligible for
such a bright star even in these narrow filters. However, as we
discuss below, the contribution to the photometric uncertainty from
the reference stars may be important. The total transit time is, as
before, approximately 55 hours, but each ring crossing (note that
there are 2 per transit) takes only 1.2 hours. This fact, coupled with
the fact that the transit depth is much larger (by a factor of $\sim
(\pi p)^{-1}$) than for a uniform disk, as shown in Figure
\ref{fg:transit_comp}, makes the narrow band transits, in principle,
much simpler to detect. Note that if one were not lucky enough to see
both ring transits, the $S/N$ would only degrade by a factor of
$\sqrt{2}$.

Photometric studies of planetary transits typically rely on precise
{\it relative} photometry in order to control systematic errors. In
this technique, the variable star's flux is measured relative to other
constant stars in the field. Observations in the narrow bands $H$ and
$K$ pose a potential problem for performing relative photometry, as
giant stars are so bright and rare that there may not be a sufficient
number of reference objects to do a reliable comparison. One way to
overcome these difficulties is to measure the Mount Wilson $S-$value,
introduced by \citet{duncan91}. The $S-$value is a self calibrating
index based on the $H$ and $K$ filters plus two broader bandpasses:
one redwards of $H$, called $R$, and one bluewards of $K$, called
$V$. Both of these reference bandpasses have a FWHM of 20\AA\ and a
rectangular instrumental profile \citep[for details
see][]{duncan91,wright04}. The $S-$value is then defined as
\begin{equation}\label{eq:svalue}
S\ =\ 2.4\ \frac{F_H\ +\ F_K}{F_R\ +\ F_V},
\end{equation}
\noindent where $F_X$ is the flux observed through band $X$.

\begin{figure}
  \begin{center}
    \plotone{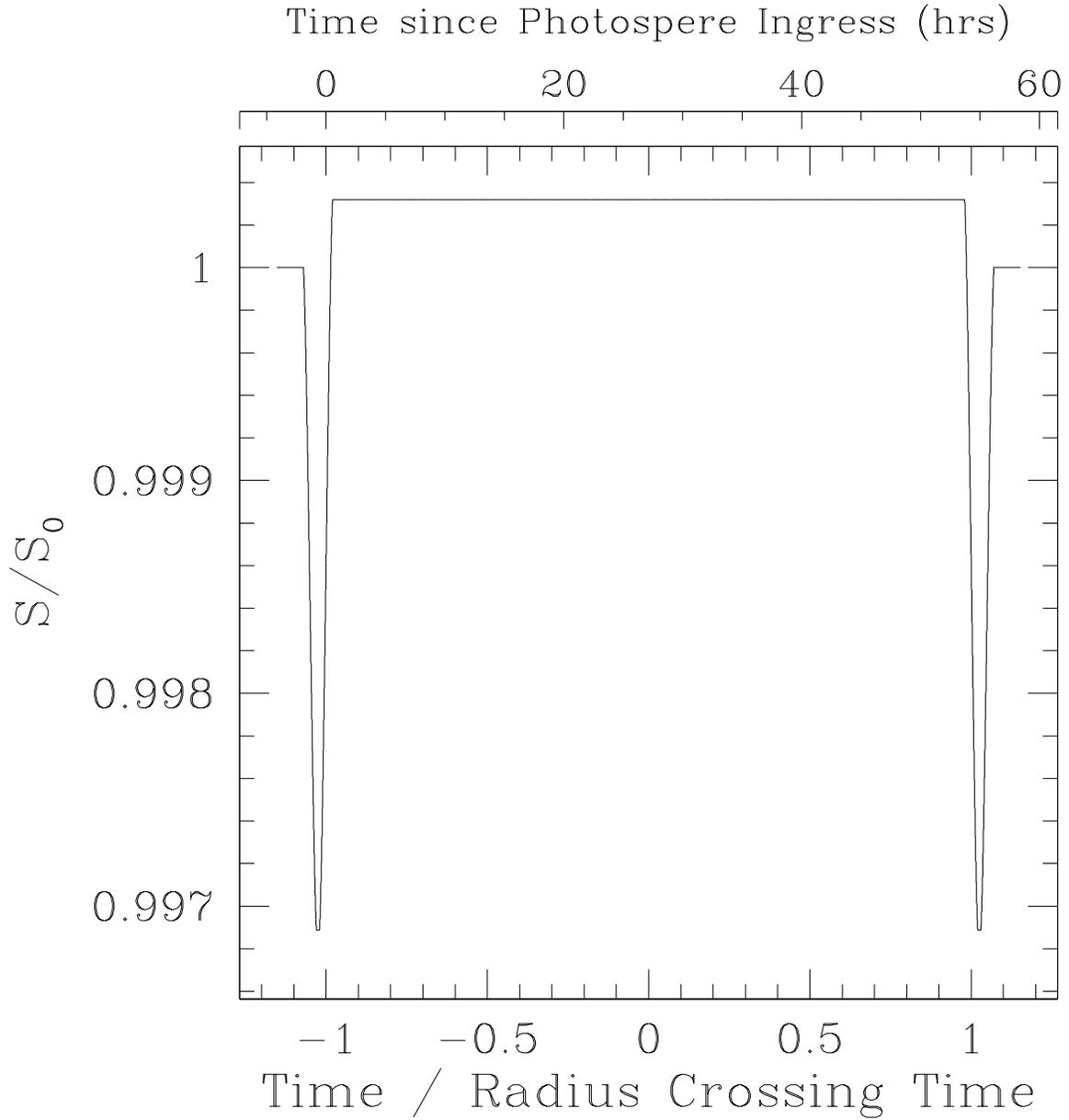}
    \caption{The light curve of a Jupiter-sized planet
    transiting a 5$R_{\odot}$ giant star, as observed
    through the Mount Wilson $S-$value.  The parameters for the system
    are described in the text and are the same as for Figure
    \ref{fg:transit_comp}.}
    \label{fg:s-transit}
  \end{center}
\end{figure}

Figure \ref{fg:s-transit} shows the shape of a transit as viewed
through the $S-$value. We have assumed the same parameters as for
Figure \ref{fg:transit_comp}. Notice that the ring transit is of equal
depth as before and that during the inner disk transit the $S-$value
is larger than outside transit, with a magnitude similar to the
broad-band transit depth. For the $S-$value, the analytic form of the
$S/N$ would be roughly the same as for the narrow band filter transits
(eqn. [\ref{eq:sn}]), as the noise is dominated by the $H$ and $K$
bands rather than by the $R$ and $V$ channels. The overall detection
$S/N$ for the $S-$value observations of the transit is $\sim 46$. If
we remove the contribution of the inner disk transit as it may be
extremely difficult to detect, the $S/N$ slightly drops to $\simeq
41$.

For estimating the signal to noise ratio, in all cases we have assumed
values of $w$ and $I$ appropriate to the 100$R_{\odot}$ star CaII
chromospheric emission model of \citet{loeb95}. These values are
unlikely to be pertinent for the fiducial $5 R_{\odot}$ giant star
host we are considering, and variations in $w$ and $I$ can affect the
detectability significantly. 

\begin{figure}
  \begin{center}
    \plotone{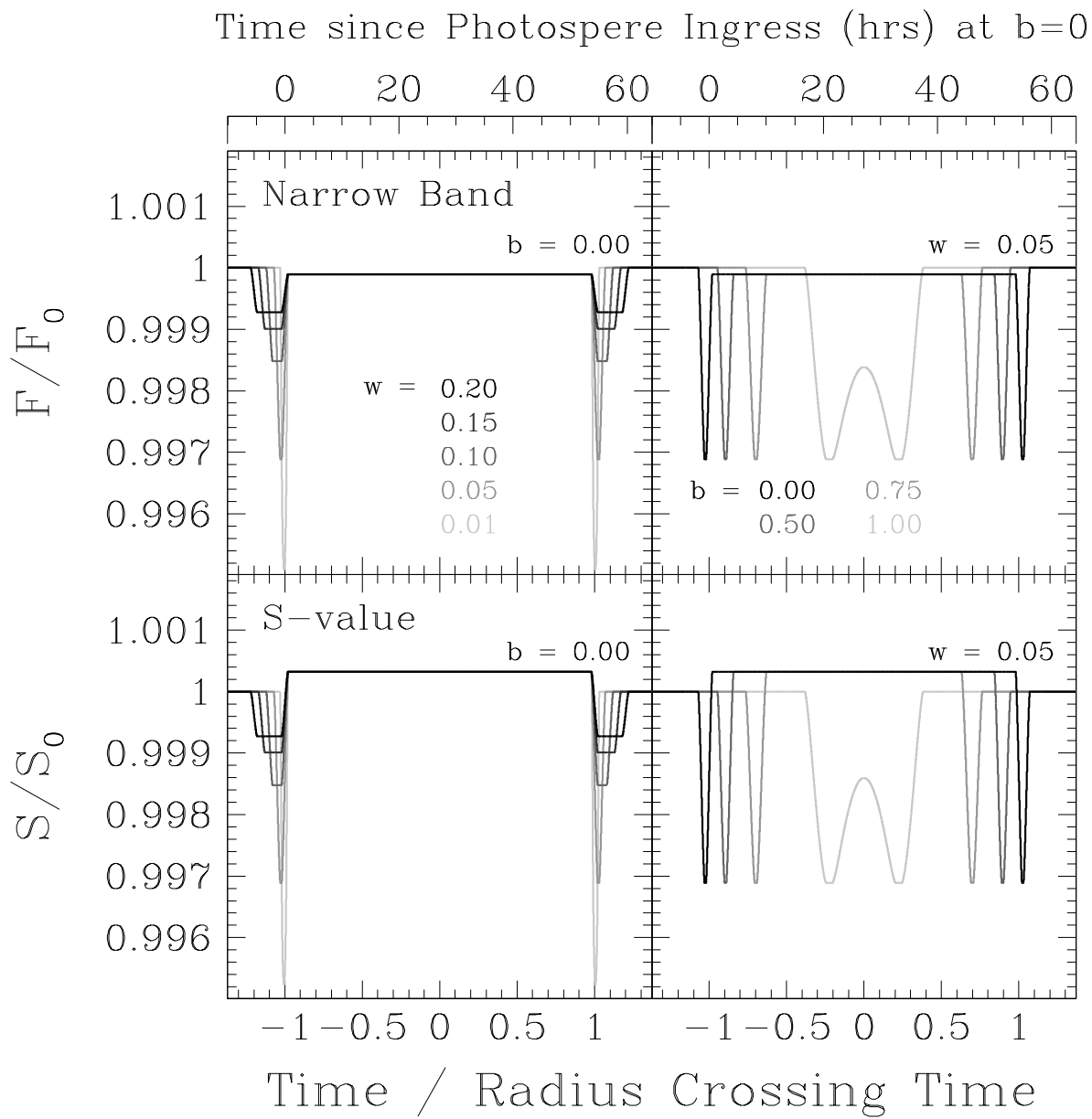}
    \caption{Transit light curves of our fiducial case for different
    values of the width of the brightened ring $w$ ({\textit{left}})
    and the impact parameter $b$ ({\it{right}}) as viewed through the
    $H$ \& $K$ narrow bands ({\it{top}}) and the Mount Wilson
    $S-$value ({\it{bottom}}). Notice that when $w$ is changed, $I$ is
    adjusted to conserve the stellar flux.}
    \label{fg:trans_change}
  \end{center}
\end{figure}

\begin{figure}
  \begin{center}
    \plotone{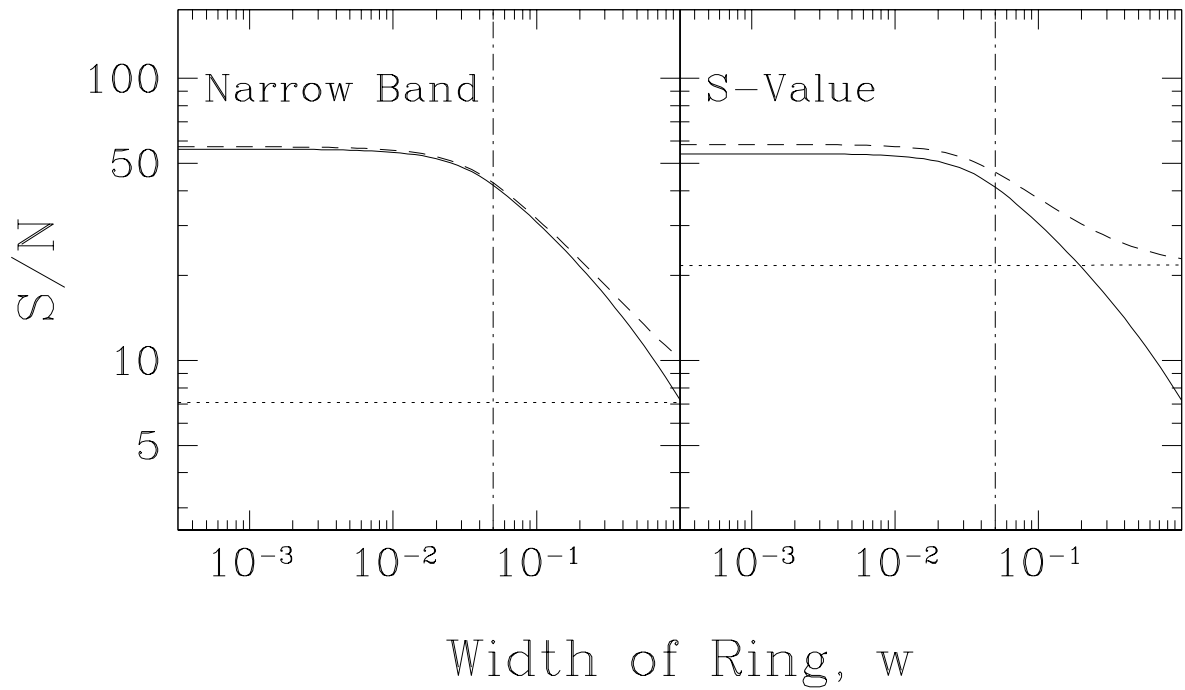}
    \caption{The solid line shows $\left(S/N\right)_{limb}$, the $S/N$
    of only the ring portion of the transit, as a function of the width
    of the ring in units of the star's radius, as observed through the
    $H$ and $K$ narrow bands ({\it{left}}), and the Mount Wilson
    $S-$value ({\it{right}}). The dashed lines show the behavior of
    the total $S/N$ (ring plus disk) while the dotted lines show the
    $S/N$ of only the disk portion of the transit. The dot-dashed line shows
    the value of $w$ we chose for the fiducial case analyzed in detail
    in the text.}
    \label{fg:SN_w}
  \end{center}
\end{figure}

Figure \ref{fg:trans_change} shows how the transit light curve of our
fiducial case changes for different effective widths of the brightened
region in the star, while Figure \ref{fg:SN_w} shows how the $S/N$
from the ring transit (i.e.\ eliminating the disk transit component of
the light curve) which we call $\left(S/N\right)_{\rm limb}$, varies
as a function of this parameter. Notice that in each case we vary $I$
to keep the total flux conserved while keeping the absolute disk
intensity fixed. In particular, notice that the $S/N$ is almost
constant for very small $w$ and then slowly decreases towards larger
values. It is unlikely that for smaller stars the width of the ring
will be larger than for our fiducial case, so our $S/N$ could be taken
as a lower limit when considering only the dependence on $w$.

\begin{figure}
  \begin{center}
    \plotone{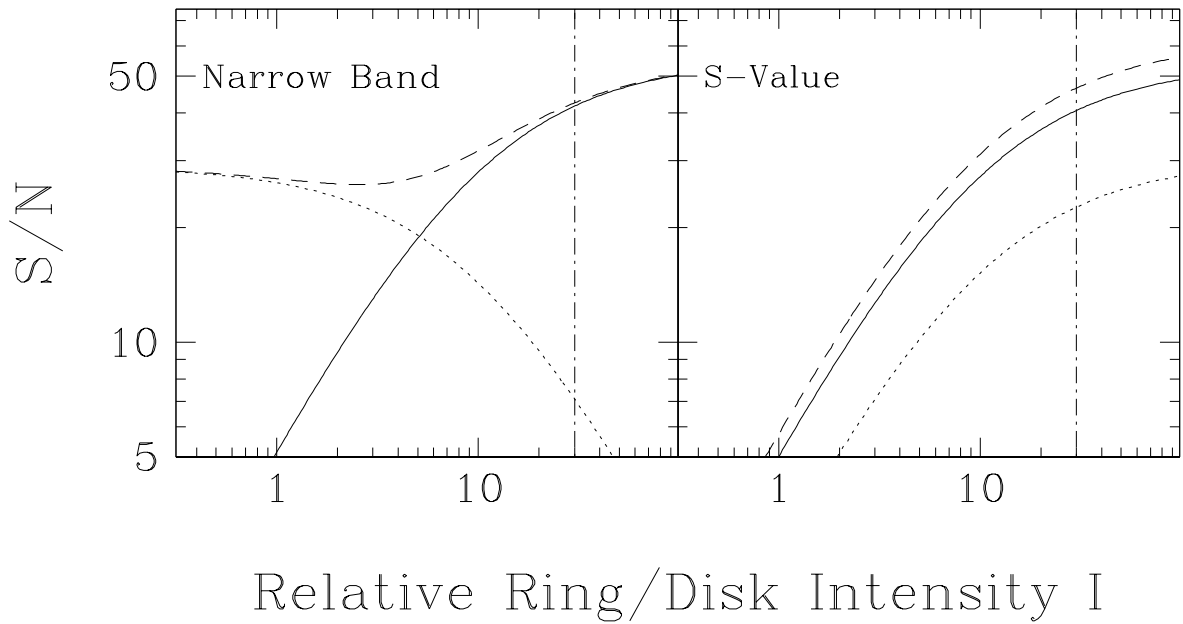}
    \caption{The solid line shows $\left(S/N\right)_{limb}$, the $S/N$
    of the only the ring part of the transit, as a function of the
    relative intensity of the ring compared to the inner disk, $I$, as
    observed through the $H$ and $K$ narrow bands ({\it{left}}), and
    the Mount Wilson $S-$value ({\it{right}}).  The dashed lines show
    the behavior of the total $S/N$ (ring plus disk) while the dotted
    lines show the $S/N$ of only the disk part of the transit. The
    dot-dashed line shows the value of $I$ we chose for the fiducial
    case analyzed in the text. Note that when $I=0$, the $S-$value is
    constant through out the transit.}
    \label{fg:SN_I}
  \end{center}
\end{figure}

If we vary $I$, the relative intensity of the ring with respect to the
disk, while keeping $w$ and the total flux fixed,
$\left(S/N\right)_{\rm limb}$ changes as shown in Figure \ref{fg:SN_I}
for the narrow band and $S-$value transits. For the latter, the
overall $S/N$ drops steadily with the value of I and is effectively
zero when $I = 0$, as all signature of the transit is erased from the
$S-$value.

\begin{figure}
  \begin{center}
    \plotone{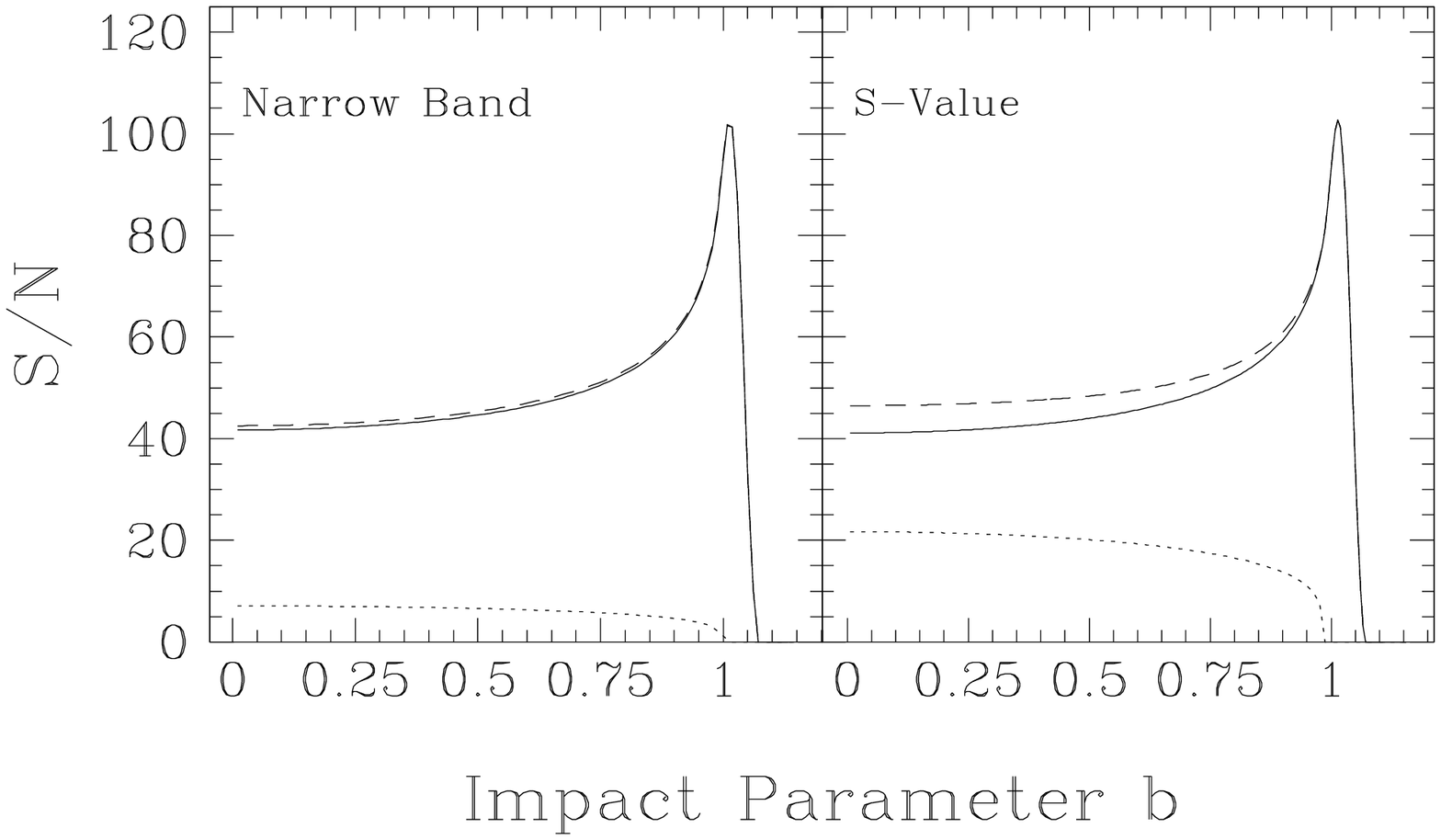}
    \caption{The solid line shows $\left(S/N\right)_{limb}$, the $S/N$
    of only the ring part of the transit, as a function of the impact
    parameter in units if the star radius, $b$, as observed through
    the $H$ and $K$ narrow bands ({\it{left}}), and the Mount Wilson
    $S-$value ({\it{right}}). The dashed lines show the behavior of
    the total $S/N$ (ring plus disk) while the dotted lines show the
    $S/N$ of only the disk part of the transit. Note that, contrary to
    broad-band transits, the $S/N$ of the detection increases as a
    function of $b$, as the planet spends more time transiting the
    limb brightened ring of the star.}
    \label{fg:SN_b}
  \end{center}
\end{figure}

We have also assumed for our fiducial case that the transit is
equatorial. Figure \ref{fg:SN_b} shows how the $S/N$ varies as a
function of the impact parameter $b$ for both narrow band and
$S-$value observations. Unlike the case for broad-band transits, the
$S/N$ increases with the impact parameter since the planet spends more
time crossing the ring. Of course, the increased duration of the ring
transit also means that some of the advantages of using these narrow
filters is lost. Figure \ref{fg:trans_change} shows how the transit
shape changes with $b$ for our fiducial case, and although for a large
range of impact parameters the duration is not significantly altered,
for a grazing transit ($b = 1$) the duration is approximately 20
hours.

While thinking of the $S-$value as a photometric index is useful for
visualizing the effect, the simplest way to measure this index is from
high resolution spectra \citep[e.g.][]{wright04}. The advantage of
this is that all four channels are observed simultaneously and longer
integration times can be achieved. For example, if we observe the
fiducial limb brightened transit described before with the HIRES
spectrograph \citep{vogt94} at the Keck observatory and measure the
$S-$value from the spectrum, we would expect the $S/N$ of the
detection to be roughly 77. To calculate this number, we have used the
HIRES $S/N$ simulator\footnote{http://www.ucolick.org/$\sim$hires/}
assuming 1.0\arcsec\ seeing and 60 second integrations and keeping all
other values to their defaults. The $S/N$ per pixel per observation is
259. Similar observations have been carried out by \citet{shkolnik03}
using the Gecko spectrograph \citep{baudrand00} at CFHT for studying
planet induced chromospheric activity in the F8.5V star HD 179949 by
using the variation in the flux of the CaII H and K lines. Also,
equivalent width variations smaller than those produced by the limb
brightened transits we have discussed here have been detected by
\citet{redfield08}, who studied NaI absorption from HD189733b using
using high resolution spectra taken with the HRS spectrograph on the
9.2 meter Hobby-Eberly Telescope. These observations suggest that the
effect we have described is currently detectable, provided the
intensity in the ring is sufficiently high relative to that of the
disk.

\subsection{Measuring the Transit Parameters}

Accurately measuring transit parameters from the light curve is
dependent on an accurate characterization of the star's CaII H \& K
surface brightness profile. This can be achieved via a priori
calculations, which are computationally expensive, or it might be
possible to constrain the surface brightness profile from the transit
light curve itself with sufficiently large $S/N$. However, two simple
limiting cases can be considered to assess the information content of
the ring transit light curves. When $w \ll p$ and when $w > 2p$, it is
possible to analytically solve for all the system parameters of
immediate interest. For simplicity, we will assume that the orbit of
the planet is circular.  We argue below that eccentric orbits do not
change our argument qualitatively.

In general, provided that $a \gg R_*$, it will be valid that
\begin{eqnarray}
P^2\ &=&\ \frac{4 \pi^2}{G M_*}\ a^3\ \ \textrm{(Kepler's
Third Law)},\\
t_P\ &=&\ \frac{P R_*}{\pi a} \ \sqrt{(1-p)^2 - b^2},
\end{eqnarray}
\noindent where $t_P$ is the time between the end of the first ring
transit and the beginning of the second one, or equivalently the time
between second and third contact with the photosphere, $v_p$ is the
speed of the planet during transit, which we have assumed to be constant,
and $b$ is the impact parameter defined as $b = (a/R_*)
\cos{i}$, where $i$ is the orbit's inclination angle. All planetary
transits must be confirmed by radial velocity measurements and, in
particular, narrow band CaII H and K transits might be easier to
observe using high resolution spectra (see \S\ref{ssec:trans_detect}),
so we will assume that the radial velocity curve of the system is
known. This yields the semi-amplitude of the radial velocity curve,
$K_*$, which is related to the orbital parameters by
\begin{equation}
K_*\ =\ \frac{2 \pi a}{P}\ \frac{M_{p}\sin{i}}{M_*},
\end{equation}
\noindent where $M_{p}$ and $M_*$ are the masses of the planet
and the star respectively. Also, from the spectra we should be able to
determine the star's surface gravity $g$,
\begin{equation}
g\ =\ \frac{G M_*}{R_*^2}.
\end{equation}
In the limit that the width of the limb brightened region is much
smaller than the radius of the planet, $w \ll p$, the depth of the ring
transit will be given by equation (\ref{eq:delta_smallw})
$$\delta_R\ =\ \frac{p}{\pi}\ \left(\frac{2 I w + \pi p/2}{2 I w +
1}\right),$$
\noindent where $\delta_R$ is the transit depth when crossing the
ring. Also, in this limit, the
time duration of one of the ring transits ($t_R/2$) is equal to the
ingress time in a typical broad-band transit, so we can state that,
\begin{equation}
\left(\frac{t_P}{t_R+t_P}\right)^2\ =\ \frac{(1-p)^2 -
b^2}{(1+p)^2-b^2}.
\end{equation}

In the other limit, when $w > 2p$, if we assume that the intensity in
the effective ring is uniform, then we know that the ring transit
depth is given by equation (\ref{eq:delta_largew}),
$$\delta_R\ =\ \frac{I p^2}{I w (w+2) + 1},$$
\noindent and that
\begin{equation}
\frac{t_R}{t_P}\ =\ \frac{\sqrt{(1+w+p)^2-b^2}}{\sqrt{(1-p)^2-b^2}} - 1.
\end{equation}
Note that in both limiting cases we have 6 equations and 7 unknowns
($a$,$M_*$,$R_*$,$p$,$b$,$M_{pl}$,$w$). The easiest
parameter to independently estimate is the radius of the star, which
can be done with broad-band colors and the calibrations of either
\citet{kervella04} or \citet{vanbelle99} plus a parallax
measurement. Given $R_*$, the system can now be solved, as we have 6
unknowns and 6 independent equations.

So far we have assumed that the orbit of the planet is perfectly
circular in order to simplify the equations. If the orbit was not
circular, we must add two more parameters to the system, the
eccentricity $e$ and the longitude of periastron $\varpi$. These two
parameters are uniquely determined by the radial velocity curve of the
star, adding no more unknowns into our system of equations.

It is unlikely that the true CaII H \& K profile of the star will
resemble exactly any of these two limits, but it is also unlikely that
it will be different enough to qualitatively change the discussion
presented here. Thus we conclude that it will be possible to estimate
the planetary parameters from a narrow band transit in the same way as
with broad-band transits. A proper determination of the typical errors
expected in the planetary parameters for a given light curve must be
calculated using accurate models of the chromosphere for such stars.

\section{Conclusions}

Studying planets around giant stars is of interest in understanding
and testing theories of planet formation and evolution. Several
planets have been found by radial velocity surveys orbiting evolved
stars, but to date none have been observed to transit. Using the data
compilation of {\it The Extrasolar Planets Encyclopedia} we have shown
that we might reasonably expect that one of the known planets orbiting
giant stars to be transiting its host as seen from
Earth. Efforts should be made in order to find and observe such an event.

We have also shown that for a typical system with a Jupiter-sized
planet and a red giant primary, the nominal $S/N$ per transit is quite
large even for a $\sim 0.5$m telescope. However, once some real-world
observational factors are taken into account, these transits may
nevertheless be quite difficult to detect. Transit durations are
extremely long ($\sim$55 hours for our fiducial system), so in
practice only a fraction of the transit would be observable from a
single observatory. Even if this is overcome by, for example,
observing from several different telescopes around the world, the
transit depth is so small that correlated red noise or intrinsic
variability could become a dominant factor of the error budget,
rendering a definitive detection impossible.  It must be kept in mind,
however, that very little is known about the behavior of the red noise
and intrinsic variability of red giants on the timescales relevant to
this problem.

Several efforts to control the systematic errors in photometric
follow-up of transits of bright dwarf stars have been carried out
quite successfully, suggesting that similar approaches might yield
positive results for transits of giant stars. Also the space transit
missions {\it Kepler} and {\it COROT} may have the photometric
precision needed to detect previously unknown transiting companions to
giant stars, provided that a sufficient number of such stars are
targeted.

While we argue that detecting broad-band planetary transits in giant
stars might be feasible in the near future, we have also described a
novel technique to detect these transits by using the CaII H \& K
chromospheric emission of red giant stars. Because the emission is
only produced in the chromosphere, at these wavelengths the light
profile of the star looks like a faint disc surrounded by a thin,
bright ring, rather than the usual broad-band picture of a
limb-darkened disc. Because the light is concentrated in a small
region of the star, a transiting planet will produce a more pronounced
feature in the star's light curve at these wavelengths.

By approximating the star by a disc surrounded by a brighter ring, we
have studied the shape of the transit light curve and the dependence
of the $S/N$ on different shapes of the profile. We have shown that,
for a typical system, transit depths are well above the red noise
amplitude of the current transit surveys searching for planets, and
that the $S/N$ of the detection can be very significant, though never
as large as the nominal value of the broad-band case. We have shown
that for two simple limiting cases, the planet radius, its mass, the
impact parameter and the semi-major axis can be inferred when the CaII
transit observations are combined with radial velocity observations
and an estimation of the star's radius.

Similar observations to those we propose here have been carried out by
other authors. These authors have in some cases achieved the precision
necessary to detect the ring transit signature of our fiducial case,
suggesting that detecting planetary transits of giants through the
variations in the CaII H \& K lines is feasible with current
instrumentation.

\acknowledgements{The authors would like to thank Jennifer Johnson and
Marc Pinsonneault for their help and suggestions, which improved the
paper significantly.  We would also like to thank Jonathan Fortney for
helpful discussions about the effects of stellar irradiation on giant
planets, and Birgit Fuhrmeister, Peter Hauschildt, Dimitar Sasselov
and Ian Short for discussions about modeling the chromospheres of
giant stars. We extend our gratitude to all the students that
participated on the discussion of a similar problem in the Order of
Magnitude course taught at The Ohio State University during the Autumn
quarter of 2006 from which this work originated. Finally, we
would also like to thank the anonymous referee for providing useful
comments and suggestions that improved this paper.}

\end{document}